\documentclass[reprint,superscriptaddress,
amsmath,amssymb, aps,prl,
]{revtex4-2}
\usepackage{graphicx}
\usepackage{amsmath,amssymb,graphicx}
\usepackage{dcolumn}
\usepackage{bm}

\begin{document}
\preprint{APS/123-QED}
\title{Improved phase locking of laser arrays with nonlinear coupling}
\author{Simon Mahler}
\email{sim.mahler@gmail.com}
\affiliation{Department of Physics of Complex Systems, Weizmann Institute of Science, Rehovot 761001, Israel}
\author{Matthew Goh}
\affiliation{Department of Physics of Complex Systems, Weizmann Institute of Science, Rehovot 761001, Israel}
\affiliation{Department of Quantum Science, RSPE, The Australian National University, Canberra ACT 0200, Australia}
\author{Chene Tradonsky}
\author{Asher A. Friesem}
\author{Nir Davidson}
\affiliation{Department of Physics of Complex Systems, Weizmann Institute of Science, Rehovot 761001, Israel}
\begin{abstract}
An arrangement based on a degenerate cavity laser for forming an array of non-linearly coupled lasers with an intra-cavity saturable absorber is presented. More than $30$ lasers were spatially phase locked and temporally Q-switched. The arrangement with nonlinear coupling was found to be $25$ times more sensitive to loss differences and converged $5$ times faster to the lowest loss phase locked state than with linear coupling, thus providing a unique solution to problems that have several near-degenerate solutions.
\end{abstract}
\maketitle
Phase locking of lasers corresponds to a state where all the lasers have the same frequency and the same constant relative phase, leading to a coherent superposition of their fields \cite{Siegman86,Glova03}. Accordingly, the total brightness of the lasers is high and allows focusing of all the lasers to a sharp spot  \cite{Kong07,Mahler19_2,Tradonsky17}. Phase locking of lasers has been incorporated in many investigations, including simulating spin systems  \cite{Nixon13,Tamate16,Berloff17}, finding the ground-state solution of complex landscapes \cite{Nixon13,Marandi14}, observing dissipative topological defects \cite{Pal17,Mahler19} and solving hard computational problems \cite{Marandi14,Tradonsky18}.

Phase locking of laser arrays can be achieved with dissipative coupling that leads to a stable state of minimal loss, which is the phase locked state \cite{Pal17,Mahler19,Nixon13}. Dissipative coupling involves mode competition whereby modes of different losses compete for the same gain \cite{Eckhouse08,Nixon13,Glova03,Marandi14}. Only modes with the lowest loss survive and are amplified by the gain medium. Accordingly, by inserting amplitude and phase linear optical elements into a laser cavity that minimize the loss of the phase locked states, it is possible to achieve phase locking with mode competition \cite{Glova03,Kong07,Mahler19_2,Zhou04}.

While phase locking with such linear optical elements has yielded many exciting results  \cite{Eckhouse08,Nixon13,Pal17,Tradonsky18,Tradonsky17,Zhou04,Mahler19_2,Guillot11,Sivaramakrishnan15,Nixon13_2}, it suffers from inherent limitations. It is very sensitive to imperfections, such as positioning errors, mechanical vibrations, thermal effects and other types of aberrations associated with these intra-cavity elements. Moreover, in many cases, especially for spin simulations and computational problem solving \cite{Marandi14,Tradonsky18,Nixon13}, there are two or more states with nearly degenerate minimal loss that cannot be distinguished from each other.

In this letter, we resort to nonlinear coupling between lasers by means of a saturable absorber (SA). A SA is a nonlinear optical element that block light until it saturates, where its optical loss decreases sharply \cite{Hercher67}. It can thus affect the temporal modes within the laser so as to obtain passive Q-switching and (longitudinal) mode locking, for generating short pulses and high output peak powers \cite{Siegman86} and studying nonlinear laser dynamics \cite{Siegman86,Kong07,Wojcik10,Winful92,Lacot96}. We show that a SA can also affect the spatial phase distribution within the laser and phase lock many individual lasers. Specifically, inserting a SA at the far-field plane of a laser array ensures that the phase locked state (that has sharp and strong intensity peaks there \cite{Tradonsky17,Nixon13,Kong07,Mahler19_2}) corresponds to the minimal loss state, to be selected by optical feedback (mode competition) \cite{Nixon13_2}.

We show experimentally and numerically that nonlinear coupling provides stable phase locking and is inherently more robust to alignment errors, aberrations and noise than linear coupling. Moreover, nonlinear mode coupling provided by the SA both in the spatial and temporal domains can yield multiple coupled copies of the laser array (corresponding to different longitudinal modes) that all converge to the same minimal loss state. Hence, significantly improving the ability of the coupled lasers to distinguish between near-degenerate states.

\begin{figure}[!ht]
\centering
\includegraphics[width=0.48\textwidth]{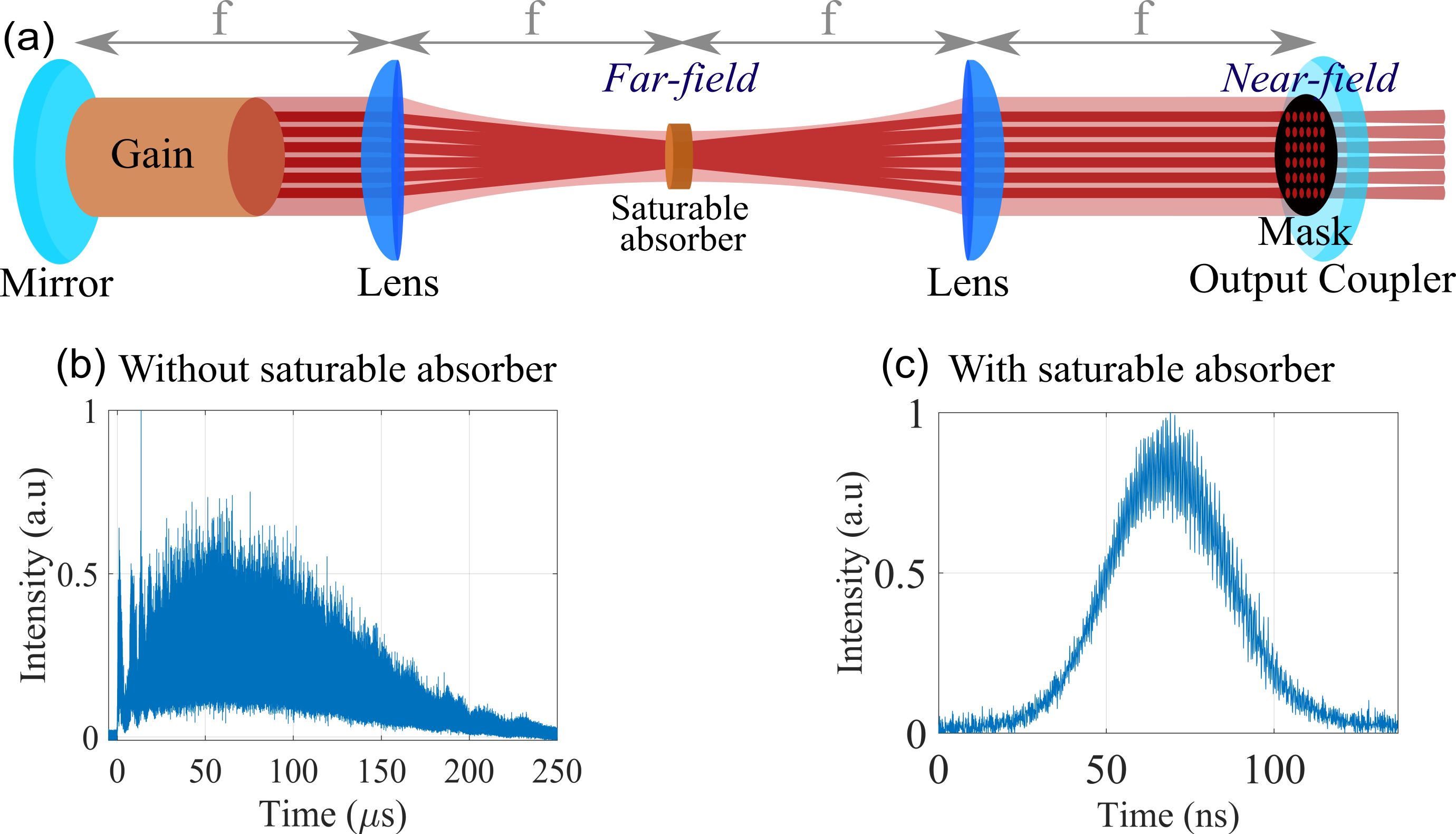}
\caption{Degenerate cavity laser (DCL) arrangement and experimental temporal evolution of the laser array intensity. (a) DCL arrangement with a mask of holes in the near-field plane so as to form an array of lasers and a saturable absorber (SA) in the far-field plane so as to non-linearly couple them. (b) Time evolution of the laser array intensity without the SA. (c) Time evolution with the SA indicating temporal Q-switching.}
\label{fig:1_exp_skecth}
\end{figure}

Our experimental arrangement with nonlinear coupling of laser arrays is based on a degenerate cavity laser (DCL) \cite{Arnaud69,Mahler19_2,Tradonsky17}, schematically presented in Fig.~\ref{fig:1_exp_skecth} (a). It was comprised of two flat mirrors where one served as a back mirror with high $99.5\%$ reflectivity and the other as an output coupler with $80\%$ reflectivity. A mask of holes for forming the array of lasers was placed at the near-field plane, adjacent to the output coupler. In our investigations, the mask was a square array of holes of diameter $200\mu m$ and period $a=300\mu m$. The gain medium was Nd:YAG rod of $0.95cm$ diameter placed adjacent to the back mirror and optically pumped by quasi-CW $100\mu s$ pulsed flash lamps operating at $1Hz$ flashing pulse rate, so operating wavelength is $\lambda=1064nm$. Between the mirrors, two spherical (Fourier) lenses of focal lengths $f=20cm$ and diameters $5.08cm$ formed a $4f$ telescope configuration. Due to the $4f$ telescope, each hole in the mask was precisely imaged onto itself after a cavity round-trip, to obtain an independent laser. 

The nonlinear coupling between the lasers was provided by a Cr:YAG saturable absorber inserted in the far-field (Fourier) plane, midway between the two lenses of the 4f telescope \cite{Mahler19_2,Tradonsky17}. For comparison, we also performed experiments with linear coupling between the lasers by displacing the output coupler by half of the Talbot length from the mask, such that the round-trip distance between them (Talbot distance) is equal to the Talbot length $Z_{T}= \frac{2a^2}{\lambda}$ \cite{Tradonsky17}.    

First, we detected the time evolution of the laser array output power without and with the SA. Figure~\ref{fig:1_exp_skecth} (b) shows the results without the SA where the lasing pulse duration was $200\mu s$ with complicated strong oscillations. Figure~\ref{fig:1_exp_skecth} (c) shows the results with the SA where the lasing pulse duration was reduced to $100ns$ indicating temporal Q-switching \cite{Siegman86}. The total energy in both cases was similar. 

\begin{figure}[!ht]
\centering
\includegraphics[width=0.48\textwidth]{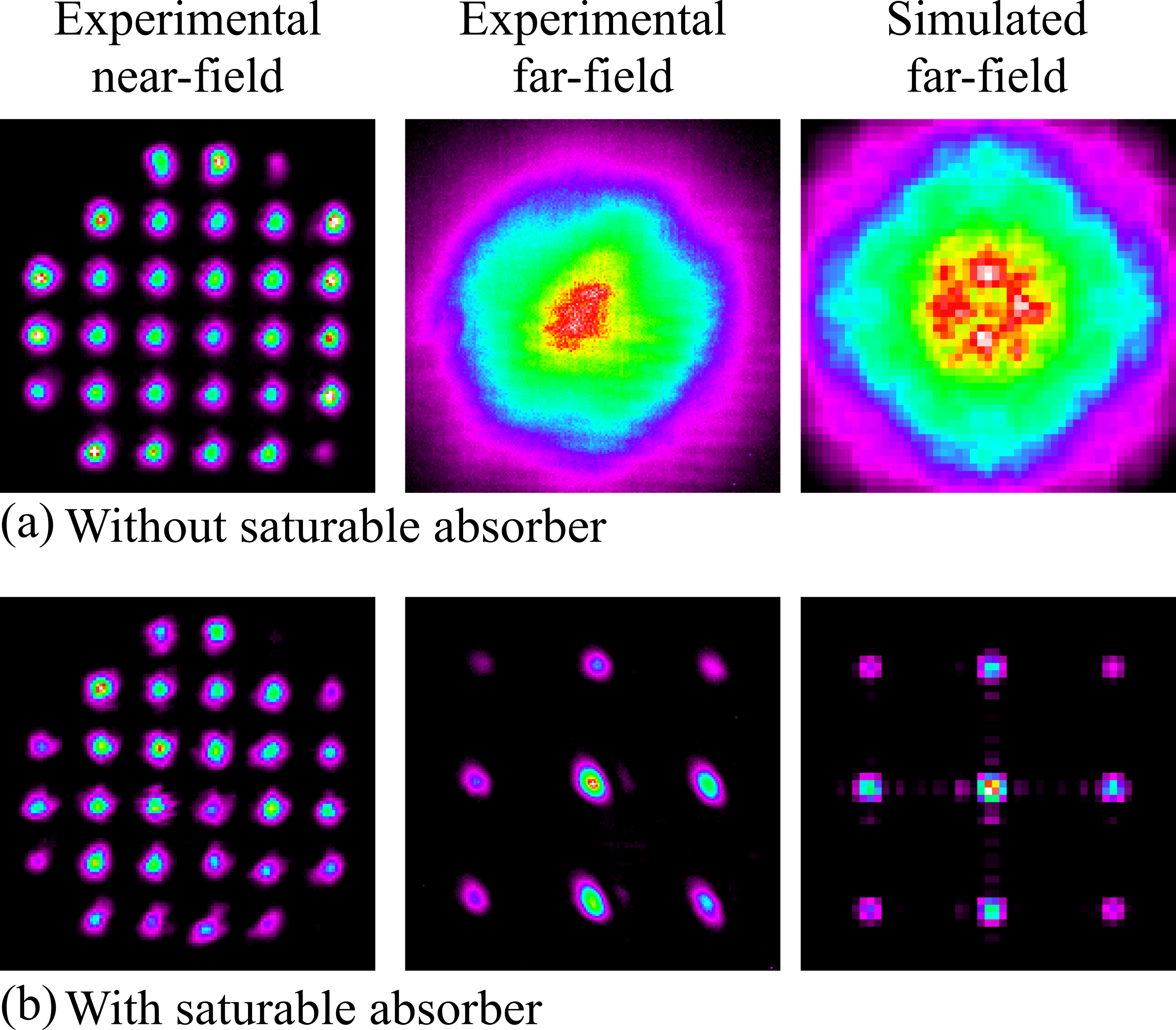}
\caption{Experimental and simulated near-field and far-field intensity distributions of the DCL arrangement. (a) Without and (b) with a SA in the far-field plane. The broad Gaussian far-field intensity distribution without the SA indicates that there is no phase relation between the lasers whereas the sharp peaks with the SA indicate near-perfect phase locking of the entire array. As evident, there is good agreement between the experimental results and the numerical simulations.} 
\label{fig:2_sa_ff_phlck}
\end{figure}

Next, we characterized the spatial coherence between the lasers by measuring the near-field and far-field intensity distributions without and with the SA. The lasers were operated close to their lasing threshold. As evident in Fig.~\ref{fig:2_sa_ff_phlck}, the near-field intensity distributions of the square array of lasers are essentially the same, while the far-field intensity distributions differ dramatically, indicating different spatial coherences. Specifically, the broad Gaussian in the far-field intensity distribution without the SA in Fig.~\ref{fig:2_sa_ff_phlck} (a) indicates no phase relation between the different lasers in the array \cite{Kong07,Tradonsky17,Mahler19_2}. 

On the other hand, the sharp peaks in the far-field intensity distribution with the SA in Fig.~\ref{fig:2_sa_ff_phlck} (b) indicate in-phase locking of most, if not all, the $30$ lasers in the array \cite{Kong07,Tradonsky17,Mahler19_2}. These high intensity peaks increase the saturation of the SA and minimize loss. This minimal nonlinear loss combined with mode competition explain the phase locking mechanism of the lasers by the SA. Even more lasers can be phase locked with the SA, but with a somewhat lower quality. We also performed numerical simulations to support our experimental results. The simulations were performed by combining the Fox-Li algorithm \cite{FoxLi61} and the Gerchberg-Saxton algorithm \cite{Gerchberg72} to obtain a combined algorithm  \cite{Tradonsky17}. As evident in Fig.~\ref{fig:2_sa_ff_phlck}, the simulated far-field intensity distributions are in good agreement with the experimental ones, indicating that the SA phase locked the lasers in the in-phase state.

\begin{figure}[!ht]
\centering
\includegraphics[width=0.48\textwidth]{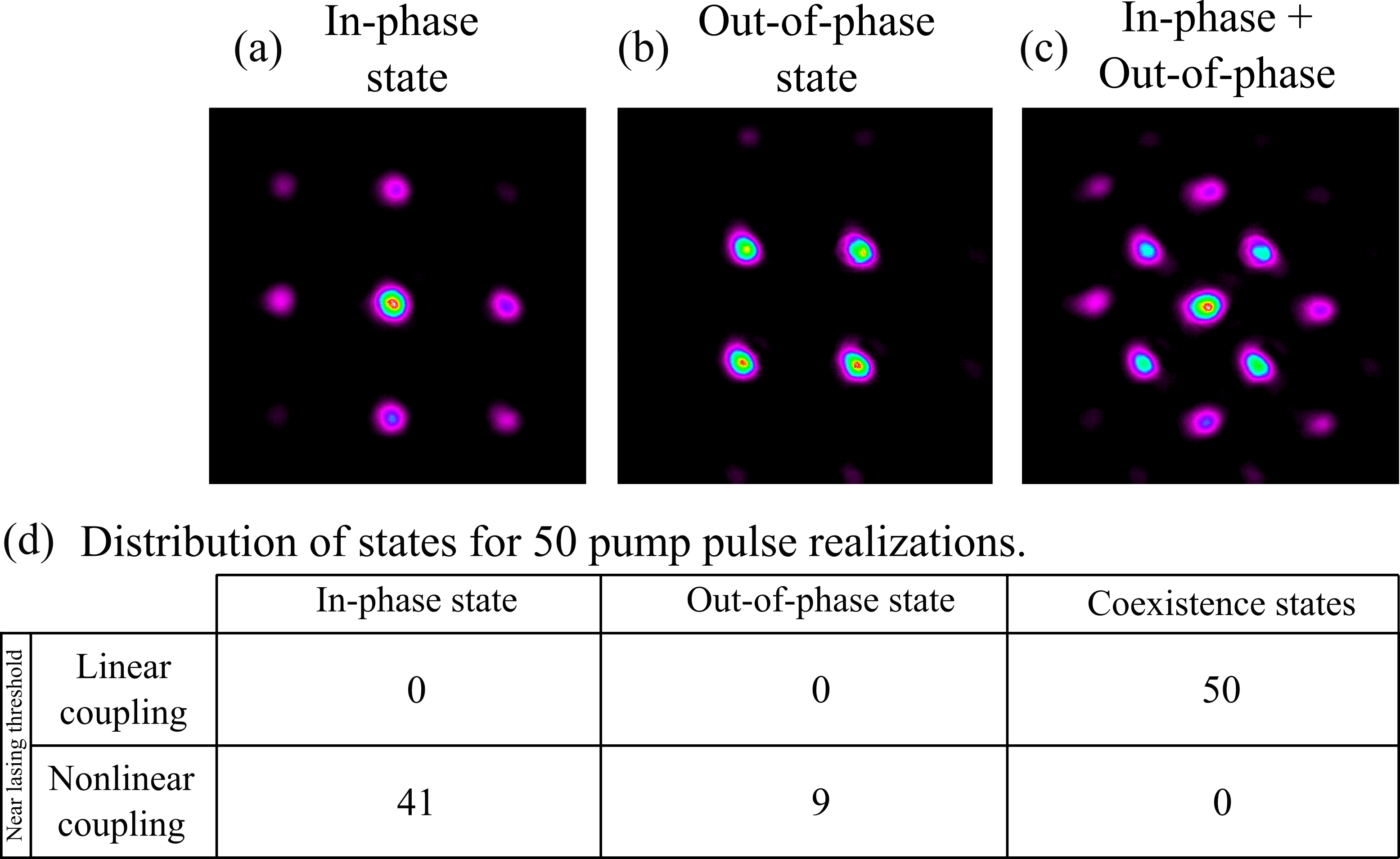}
\caption{Typical experimental far-field intensity distributions for different pump pulse realizations. (a) In-phase state with SA, (b) out-of-phase state with SA and (c) coexistence state of in-phase and out-of-phase states with SA. (d) Distribution of the number of states for $50$ different pump pulse realizations, top row: linear coupling (Talbot diffraction), bottom row: nonlinear coupling (SA) near lasing threshold.}
\label{fig:3_sa_divers_phlck}
\end{figure}

For some pump pulse realizations, other phase locked states such as the out-of-phase state \cite{Kong07,Tradonsky17,Mahler19_2} can occur. The other phase locked states also have sharp peaks in their far-field intensity distribution and similarly minimize loss. Some typical results are presented in Fig.~\ref{fig:3_sa_divers_phlck}, showing experimental far-field intensity distributions with a SA for different pump pulse realizations. Figures~\ref{fig:3_sa_divers_phlck} (a) and (b) show the in-phase and out-of-phase states, and Fig.~\ref{fig:3_sa_divers_phlck} (c) shows a coexistence state of the in-phase and out-of-phase states. To explain such a coexistence state, we note that each laser in the array contains several hundreds of temporal (longitudinal) modes \cite{Nixon13,Tradonsky17}. Each of these longitudinal modes corresponds to a different realization of the spatially coupled lasers so phase locking could be either in the in-phase state or in the out-of-phase state. In the coexistence state, part of the longitudinal modes phase locked in-phase and part out-of-phase \cite{Nixon13,Tradonsky17}.

We determined the likelihood of the coexistence state when using either linear or nonlinear coupling. The results are presented in Fig.~\ref{fig:3_sa_divers_phlck} (d), for $50$ different pump pulse realizations where all the near-field intensity distributions were the same. For the linear coupling (by Talbot diffraction with Talbot distance $Z_{T}$ for which the two phase locked states are exactly degenerated), we always observed coexistence states only (top row). Such behavior can be easily understood: for linear coupling, the different longitudinal modes are uncoupled and act as an ensemble of hundreds of independent realizations, where each randomly selects a different phase locked state. The probability that all of them select the same state is exponentially small. For linear coupling, the loss of the coexistence state is minimal, similar to that of in-phase or out-of-phase state. 

For nonlinear coupling (with SA), we found that the likelihood of the coexistence state is completely suppressed near the lasing threshold and all the longitudinal modes choose the same phase locked state (bottom row). Such suppression can be explained by noting that in the coexistence state, there are many far-field peaks whose intensity is relatively low and saturate less the SA, thereby increasing the loss. In addition, the nonlinear coupling between longitudinal modes provided by the SA forces the longitudinal modes to have the same phase locked state. As a result, near lasing threshold where mode competition is the strongest, a single phase locked state is enforced and the coexistence state is suppressed \cite{Note0}.
We also found that high above the lasing threshold, the likelihood of the coexistence state was not completely suppressed although it is lower with nonlinear coupling than with linear.

\begin{figure}[!ht]
\centering
\includegraphics[width=0.47\textwidth]{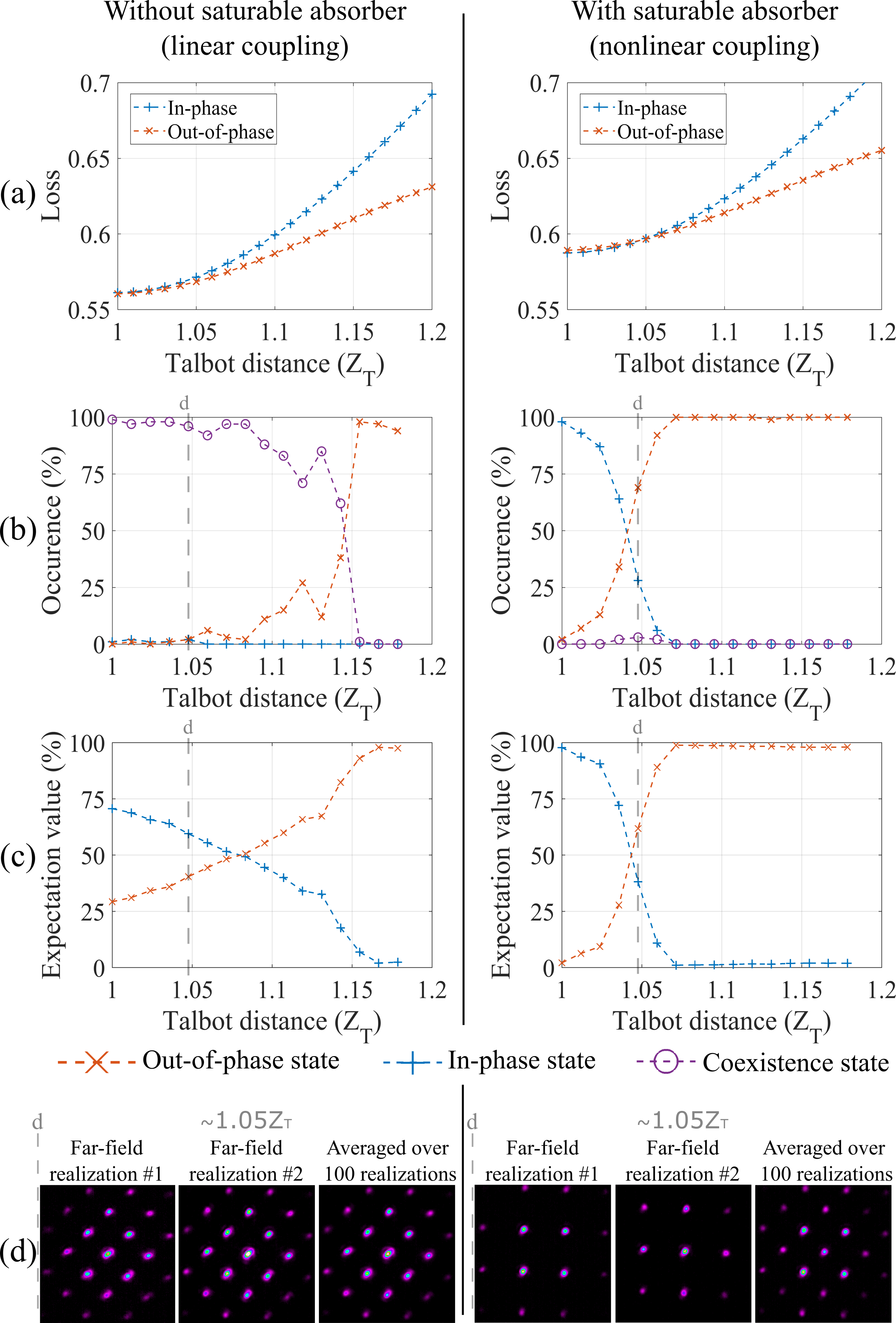}
\caption{The effect of the SA on the convergence to the lowest loss phase locked state. (a) Calculated losses of the in-phase and out-of-phase states as a function of Talbot distance in the range $[1~\mathrm{to}~1.2]Z_{T}$. The SA increases the loss of the out-of-phase state by $\sim0.5\%$ as compared to the in-phase state thereby shifting the degeneracy distance between them from $1Z_{T}$. to $1.05Z_{T}$. (b) Measured relative occurrence of the in-phase, out-of-phase and coexistence states as a function of Talbot distance. (c) Measured expectation value of the in-phase and out-of-phase states as a function of Talbot distance. (d) Typical far-field intensity distribution for single realizations and averaged distribution over the $100$ realizations at $1.05Z_{T}$.}
\label{fig:4_better_solver_graphs}
\end{figure}

Next, we investigated whether adding nonlinear coupling to linearly coupled lasers can improve the convergence to the lowest loss phase locked state (e.g. out-of-phase state), when an additional state (e.g. in-phase state) with nearly identical but slightly higher loss is present. The effects of the SA on the convergence to the lowest loss phase locked state are presented in Fig.~\ref{fig:4_better_solver_graphs}. The linear losses of the in-phase and out-of-phase states were controlled by varying the Talbot distance (round-trip distance between the near-field mask and the output coupler) \cite{Tradonsky17}. 

Using an eigenvalues modal analysis \cite{Mehuys91}, we calculated the losses, without and with a SA, of the in-phase and out-of-phase states for different Talbot distances in the range $[1~\mathrm{to}~1.2]Z_{T}$. The results are presented in Fig.~\ref{fig:4_better_solver_graphs} (a). For linear coupling, there is a degeneracy between the two phase locked states at $1Z_{T}$. The difference in loss between the two states increases as the Talbot distance increases, where the loss of the out-of-phase state becomes significantly lower \cite{Note1}. When adding nonlinear coupling, the calculated losses of the in-phase state are slightly reduced (by $\sim0.5\%$) compared with those of the out-of-phase state for all Talbot distances, shifting the degeneracy distance from $1Z_{T}$ to about $1.04Z_{T}$ (see Fig.~\ref{fig:4_better_solver_graphs} (a) right) \cite{Note2}.

We measured the relative occurrence of each phase locked state (in-phase, out-of-phase or coexistence state) as a function of the Talbot distance for $100$ different pump pulse realizations \cite{Note3}. The results are presented in Fig.~\ref{fig:4_better_solver_graphs} (b). For linear coupling, the minimal loss phase locked state (i.e. out-of-phase state) with $\geq90\%$ occurrence rate is achieved only at distances larger than $1.15Z_{T}$ where the loss difference between in-phase and out-of-phase states is large ($\simeq5\%$ in Fig.~\ref{fig:4_better_solver_graphs} (a) left). For smaller loss difference, the coexistence state is dominant, indicating that some of the longitudinal modes select the wrong phase locked state.

For the nonlinear coupling, the occurrence of the coexistence state is negligible (occurs only around the degeneracy distance $1.04Z_{T}$ with a very small probability), confirming that with nonlinear coupling all longitudinal modes select the same phase locked state (as already noted in Fig.~\ref{fig:3_sa_divers_phlck} without Talbot diffraction). As the Talbot distance is varied across the degeneracy distance $1.04Z_{T}$, a sharp transition occurs between the in-phase and the out-of-phase states. The minimal loss state with $\geq90\%$ occurrence rate is achieved at Talbot distances $1.03Z_{T}$ and $1.06Z_{T}$ when the difference in loss is $\simeq0.2\%$ ($\simeq25$ times smaller than for linear coupling). We also found that the results for the nonlinear coupling are more robust and repeatable.

Figure~\ref{fig:4_better_solver_graphs} (c) shows the expectation value of the in-phase and out-of-phase states as a function of the Talbot distance (obtained by summing the measured far-field intensity distributions of all $100$ realizations and calculating the expectation intensity value of each phase locked state). For nonlinear coupling, the results in Figs.~\ref{fig:4_better_solver_graphs} (b) and (c) have the same behavior (simply because there are only in-phase or out-of-phase states with no coexistence state). For linear coupling, the minimal loss state with $\geq90\%$ expectation value is achieved again only for Talbot distances larger than $1.15Z_{T}$, where the difference in loss is large ($\sim5\%$). Even ``majority selection'' of the longitudinal modes (expectation value above $50\%$) fails for Talbot distances smaller than $1.08Z_{T}$, where the difference in loss is $\simeq1.3\%$. The improvement of nonlinear coupling over linear coupling for finding the lowest loss state can also be quantified by using the slope of the transition from the out-of-phase to in-phase state, which is $\simeq5$ times sharper with nonlinear coupling.

Finally, Fig.~\ref{fig:4_better_solver_graphs} (d) shows experimental far-field intensity distributions at $1.05Z_{T}$ for two typical realizations and the averaged distribution over the $100$ realizations, for both linear and nonlinear coupling. For linear coupling, the distributions for all the different single realizations are identical (and hence also is the averaged distribution) and correspond to the coexistence state, indicating that the longitudinal modes are uncoupled. For nonlinear coupling, the distribution for the single realizations correspond either to the in-phase or out-of-phase state (no coexistence state), indicating that the longitudinal modes are coupled and the averaged distribution reveals both states. 

We showed that nonlinear coupling with SA couples many longitudinal modes so they all phase lock to the same state. We also showed that nonlinear coupling significantly improves the ability of an array of lasers to find the correct minimal loss state, analogous to mapping the ground state of the classical XY spin Hamiltonian \cite{Nixon13}. To support this analogy, we offer a toy model of coupled Ising spins in a magnetic field and show that its properties are analogous to those of our coupled longitudinal modes. 

\begin{figure}[!ht]
\centering
\includegraphics[width=0.27\textwidth]{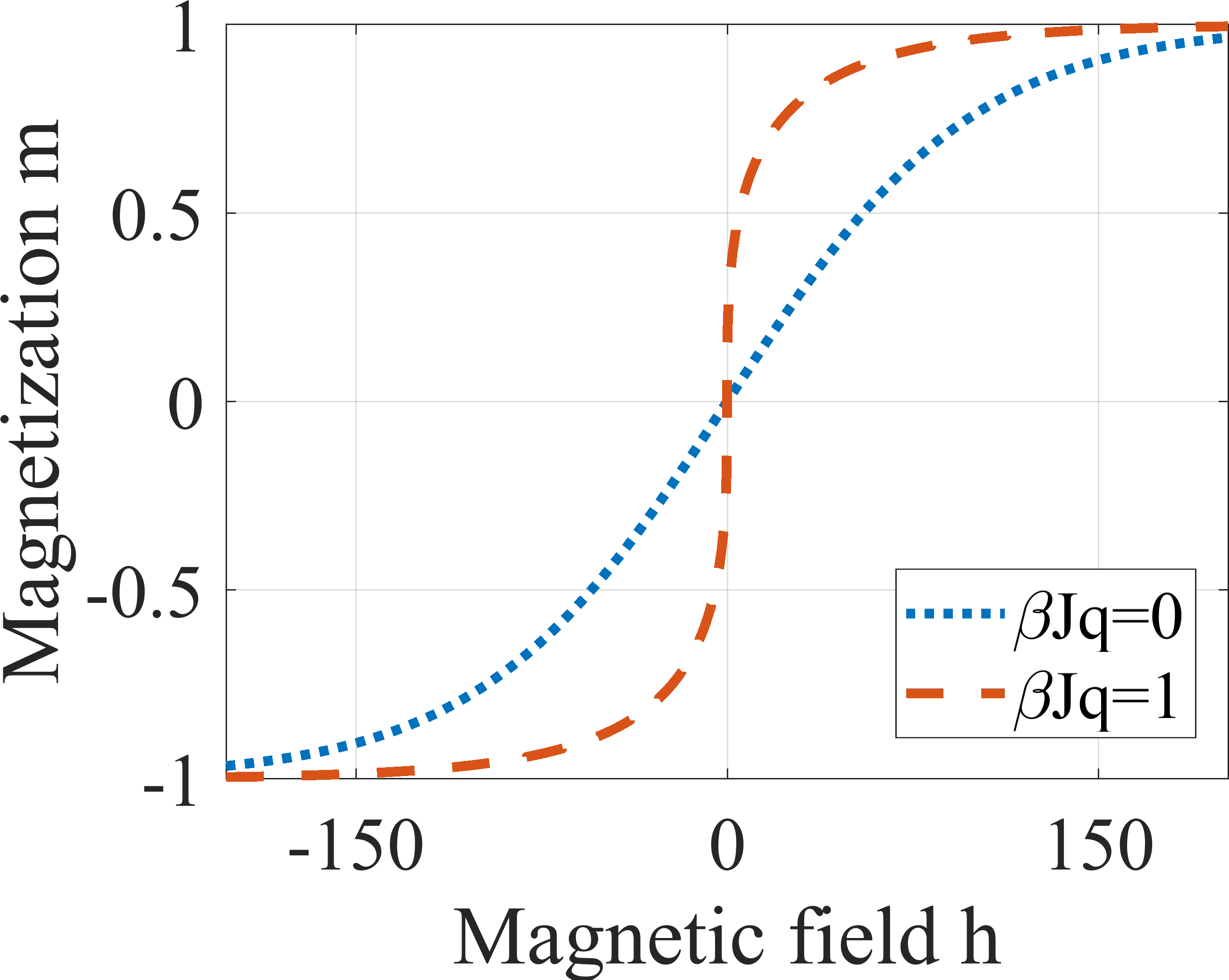}
\caption{Toy model numerical results. The evolution of the magnetization of the spins m as a function of the magnetic field h for the spins of an Ising model under mean-field approximation with fixed $\beta=0.01$ and two different coupling strengths (zero) $\beta Jq=0$ and (high) $\beta Jq=1$. The sharp (soft) transition of the magnetization around the energy-degeneracy point $h=0$ at zero (high) coupling is analogous to the sharp (soft) transition between the two phase locked states of the non-linearly (linearly) coupled lasers in Fig.~\ref{fig:4_better_solver_graphs} (c).}
\label{fig:5_magnetization_profile}
\end{figure}

Consider $N$ Ising spins in a magnetic field $h$ that need to correctly find the minimal energy state of the Ising Hamiltonian, described by the Ising model \cite{Stanley71,Strecka05}. In the mean field approximation, the magnetization $m$ of the spins, that determine their alignment, is \cite{Stanley71,Strecka05}:
\begin{equation}
m=tanh(\beta(Jqm + h)),
\label{eq:spin_m}
\end{equation}
where $q$ is the coordination number, $J$ is the coupling between spins and $\beta=\frac{1}{k_{B}T}$ is the inverse temperature. The magnetization of the spins as a function of the magnetic field is analogous to the coherence of the lasers (phase locking) where the magnetic field is analogous to the loss difference. We numerically solved Eq.~\ref{eq:spin_m} at fixed $\beta$ and $q$, to find the magnetization $m$ as a function of the magnetic field $h$, for two different coupling strengths (zero) $\beta Jq=0$ and (high) $\beta Jq=1$ (see  Fig.~\ref{fig:5_magnetization_profile}). 

As evident, the uncoupled spins (analogous to the uncoupled longitudinal modes in our lasers) produce significant magnetization only when $h>1/\beta$, as for a single spin. However, the strongly coupled spins (analogous to the strongly coupled longitudinal modes) align with the magnetic field and produce significant magnetization even for $h<<1/\beta$ and are thus much more sensitive to the direction of $h$ than the uncoupled spins.

To conclude, we showed that an array of $30$ lasers can be phase locked efficiently and robustly, by resorting to nonlinear coupling with a saturable absorber in the far-field. The nonlinear coupling was found to significantly improve the ability of the lasers to converge to the correct minimal loss phase locked state, that is mapped to the ground state of the classical XY spin Hamiltonian \cite{Nixon13}. It is $25$ times more sensitive to differences in loss with $5$ times faster convergence to the lowest loss state than with linear coupling. The nonlinear coupling forces all longitudinal modes of the lasers to have the same phase locked state (in analogy to strongly coupled Ising spins that produce significant magnetization even for magnetic energy much smaller than their thermal energy). We expect that nonlinear coupling can be extended for solving hard computational problems that have many near-degenerate solutions \cite{Nixon13}.

\begin{acknowledgments}
The authors wish to acknowledge the Israel Science Foundation and the Israeli Planning and Budgeting Committee Fellowship Program for their support.
\end{acknowledgments}

\bibliographystyle{apsrev4-1}
\bibliography{apssamp}

\begin{thebibliography}{31}%
\makeatletter
\providecommand \@ifxundefined [1]{%
 \@ifx{#1\undefined}
}%
\providecommand \@ifnum [1]{%
 \ifnum #1\expandafter \@firstoftwo
 \else \expandafter \@secondoftwo
 \fi
}%
\providecommand \@ifx [1]{%
 \ifx #1\expandafter \@firstoftwo
 \else \expandafter \@secondoftwo
 \fi
}%
\providecommand \natexlab [1]{#1}%
\providecommand \enquote  [1]{``#1''}%
\providecommand \bibnamefont  [1]{#1}%
\providecommand \bibfnamefont [1]{#1}%
\providecommand \citenamefont [1]{#1}%
\providecommand \href@noop [0]{\@secondoftwo}%
\providecommand \href [0]{\begingroup \@sanitize@url \@href}%
\providecommand \@href[1]{\@@startlink{#1}\@@href}%
\providecommand \@@href[1]{\endgroup#1\@@endlink}%
\providecommand \@sanitize@url [0]{\catcode `\\12\catcode `\$12\catcode
  `\&12\catcode `\#12\catcode `\^12\catcode `\_12\catcode `\%12\relax}%
\providecommand \@@startlink[1]{}%
\providecommand \@@endlink[0]{}%
\providecommand \url  [0]{\begingroup\@sanitize@url \@url }%
\providecommand \@url [1]{\endgroup\@href {#1}{\urlprefix }}%
\providecommand \urlprefix  [0]{URL }%
\providecommand \Eprint [0]{\href }%
\providecommand \doibase [0]{http://dx.doi.org/}%
\providecommand \selectlanguage [0]{\@gobble}%
\providecommand \bibinfo  [0]{\@secondoftwo}%
\providecommand \bibfield  [0]{\@secondoftwo}%
\providecommand \translation [1]{[#1]}%
\providecommand \BibitemOpen [0]{}%
\providecommand \bibitemStop [0]{}%
\providecommand \bibitemNoStop [0]{.\EOS\space}%
\providecommand \EOS [0]{\spacefactor3000\relax}%
\providecommand \BibitemShut  [1]{\csname bibitem#1\endcsname}%
\let\auto@bib@innerbib\@empty
\bibitem [{\citenamefont {Siegman}(1986)}]{Siegman86}%
  \BibitemOpen
  \bibfield  {author} {\bibinfo {author} {\bibfnamefont {A.~E.}\ \bibnamefont
  {Siegman}},\ }\href@noop {} {\emph {\bibinfo {title} {Lasers}}},\ \bibinfo
  {edition} {1st}\ ed.,\ \bibinfo {series} {1}, Vol.~\bibinfo {volume} {1}\
  (\bibinfo  {publisher} {University Science Books},\ \bibinfo {address} {Mill
  Valey, California},\ \bibinfo {year} {1986})\BibitemShut {NoStop}%
\bibitem [{\citenamefont {Glova}(2003)}]{Glova03}%
  \BibitemOpen
  \bibfield  {author} {\bibinfo {author} {\bibfnamefont {A.~F.}\ \bibnamefont
  {Glova}},\ }\href {\doibase 10.1070/qe2003v033n04abeh002415} {\bibfield
  {journal} {\bibinfo  {journal} {Quantum Electronics}\ }\textbf {\bibinfo
  {volume} {33}},\ \bibinfo {pages} {283} (\bibinfo {year} {2003})}\BibitemShut
  {NoStop}%
\bibitem [{\citenamefont {Kong}\ \emph {et~al.}(2007)\citenamefont {Kong},
  \citenamefont {Liu}, \citenamefont {Sanders}, \citenamefont {Chen},\ and\
  \citenamefont {Lee}}]{Kong07}%
  \BibitemOpen
  \bibfield  {author} {\bibinfo {author} {\bibfnamefont {F.}~\bibnamefont
  {Kong}}, \bibinfo {author} {\bibfnamefont {L.}~\bibnamefont {Liu}}, \bibinfo
  {author} {\bibfnamefont {C.}~\bibnamefont {Sanders}}, \bibinfo {author}
  {\bibfnamefont {Y.~C.}\ \bibnamefont {Chen}}, \ and\ \bibinfo {author}
  {\bibfnamefont {K.~K.}\ \bibnamefont {Lee}},\ }\href {\doibase
  10.1063/1.2721390} {\bibfield  {journal} {\bibinfo  {journal} {Applied
  Physics Letters}\ }\textbf {\bibinfo {volume} {90}},\ \bibinfo {pages}
  {151110} (\bibinfo {year} {2007})}\BibitemShut {NoStop}%
\bibitem [{\citenamefont {Mahler}\ \emph
  {et~al.}(2019{\natexlab{a}})\citenamefont {Mahler}, \citenamefont
  {Tradonsky}, \citenamefont {Chriki}, \citenamefont {Friesem},\ and\
  \citenamefont {Davidson}}]{Mahler19_2}%
  \BibitemOpen
  \bibfield  {author} {\bibinfo {author} {\bibfnamefont {S.}~\bibnamefont
  {Mahler}}, \bibinfo {author} {\bibfnamefont {C.}~\bibnamefont {Tradonsky}},
  \bibinfo {author} {\bibfnamefont {R.}~\bibnamefont {Chriki}}, \bibinfo
  {author} {\bibfnamefont {A.~A.}\ \bibnamefont {Friesem}}, \ and\ \bibinfo
  {author} {\bibfnamefont {N.}~\bibnamefont {Davidson}},\ }\href {\doibase
  10.1364/OSAC.2.002077} {\bibfield  {journal} {\bibinfo  {journal} {OSA
  Continuum}\ }\textbf {\bibinfo {volume} {2}},\ \bibinfo {pages} {2077}
  (\bibinfo {year} {2019}{\natexlab{a}})}\BibitemShut {NoStop}%
\bibitem [{\citenamefont {Tradonsky}\ \emph {et~al.}(2017)\citenamefont
  {Tradonsky}, \citenamefont {Pal}, \citenamefont {Chriki}, \citenamefont
  {Davidson},\ and\ \citenamefont {Friesem}}]{Tradonsky17}%
  \BibitemOpen
  \bibfield  {author} {\bibinfo {author} {\bibfnamefont {C.}~\bibnamefont
  {Tradonsky}}, \bibinfo {author} {\bibfnamefont {V.}~\bibnamefont {Pal}},
  \bibinfo {author} {\bibfnamefont {R.}~\bibnamefont {Chriki}}, \bibinfo
  {author} {\bibfnamefont {N.}~\bibnamefont {Davidson}}, \ and\ \bibinfo
  {author} {\bibfnamefont {A.~A.}\ \bibnamefont {Friesem}},\ }\href {\doibase
  10.1364/AO.56.00A126} {\bibfield  {journal} {\bibinfo  {journal} {Appl.
  Opt.}\ }\textbf {\bibinfo {volume} {56}},\ \bibinfo {pages} {A126} (\bibinfo
  {year} {2017})}\BibitemShut {NoStop}%
\bibitem [{\citenamefont {Nixon}\ \emph
  {et~al.}(2013{\natexlab{a}})\citenamefont {Nixon}, \citenamefont {Ronen},
  \citenamefont {Friesem},\ and\ \citenamefont {Davidson}}]{Nixon13}%
  \BibitemOpen
  \bibfield  {author} {\bibinfo {author} {\bibfnamefont {M.}~\bibnamefont
  {Nixon}}, \bibinfo {author} {\bibfnamefont {E.}~\bibnamefont {Ronen}},
  \bibinfo {author} {\bibfnamefont {A.~A.}\ \bibnamefont {Friesem}}, \ and\
  \bibinfo {author} {\bibfnamefont {N.}~\bibnamefont {Davidson}},\ }\href
  {\doibase 10.1103/PhysRevLett.110.184102} {\bibfield  {journal} {\bibinfo
  {journal} {Phys. Rev. Lett.}\ }\textbf {\bibinfo {volume} {110}},\ \bibinfo
  {pages} {184102} (\bibinfo {year} {2013}{\natexlab{a}})}\BibitemShut
  {NoStop}%
\bibitem [{\citenamefont {Tamate}\ \emph {et~al.}(2016)\citenamefont {Tamate},
  \citenamefont {Yamamoto}, \citenamefont {Marandi}, \citenamefont {McMahon},\
  and\ \citenamefont {Utsunomiya}}]{Tamate16}%
  \BibitemOpen
  \bibfield  {author} {\bibinfo {author} {\bibfnamefont {S.}~\bibnamefont
  {Tamate}}, \bibinfo {author} {\bibfnamefont {Y.}~\bibnamefont {Yamamoto}},
  \bibinfo {author} {\bibfnamefont {A.}~\bibnamefont {Marandi}}, \bibinfo
  {author} {\bibfnamefont {P.}~\bibnamefont {McMahon}}, \ and\ \bibinfo
  {author} {\bibfnamefont {S.}~\bibnamefont {Utsunomiya}},\ }\href@noop {}
  {\enquote {\bibinfo {title} {Simulating the classical {XY} model with a laser
  network},}\ } (\bibinfo {year} {2016}),\ \Eprint
  {http://arxiv.org/abs/arXiv:1608.00358} {arXiv:1608.00358} \BibitemShut
  {NoStop}%
\bibitem [{\citenamefont {Berloff}\ \emph {et~al.}(2017)\citenamefont
  {Berloff}, \citenamefont {Silva}, \citenamefont {Kalinin}, \citenamefont
  {Askitopoulos}, \citenamefont {T{\"o}pfer}, \citenamefont {Cilibrizzi},
  \citenamefont {Langbein},\ and\ \citenamefont {Lagoudakis}}]{Berloff17}%
  \BibitemOpen
  \bibfield  {author} {\bibinfo {author} {\bibfnamefont {N.~G.}\ \bibnamefont
  {Berloff}}, \bibinfo {author} {\bibfnamefont {M.}~\bibnamefont {Silva}},
  \bibinfo {author} {\bibfnamefont {K.}~\bibnamefont {Kalinin}}, \bibinfo
  {author} {\bibfnamefont {A.}~\bibnamefont {Askitopoulos}}, \bibinfo {author}
  {\bibfnamefont {J.~D.}\ \bibnamefont {T{\"o}pfer}}, \bibinfo {author}
  {\bibfnamefont {P.}~\bibnamefont {Cilibrizzi}}, \bibinfo {author}
  {\bibfnamefont {W.}~\bibnamefont {Langbein}}, \ and\ \bibinfo {author}
  {\bibfnamefont {P.~G.}\ \bibnamefont {Lagoudakis}},\ }\href
  {https://doi.org/10.1038/nmat4971} {\bibfield  {journal} {\bibinfo  {journal}
  {Nature Materials}\ }\textbf {\bibinfo {volume} {16}},\ \bibinfo {pages}
  {1120 EP } (\bibinfo {year} {2017})},\ \bibinfo {note} {article}\BibitemShut
  {NoStop}%
\bibitem [{\citenamefont {Marandi}\ \emph {et~al.}(2014)\citenamefont
  {Marandi}, \citenamefont {Wang}, \citenamefont {Takata}, \citenamefont
  {Byer},\ and\ \citenamefont {Yamamoto}}]{Marandi14}%
  \BibitemOpen
  \bibfield  {author} {\bibinfo {author} {\bibfnamefont {A.}~\bibnamefont
  {Marandi}}, \bibinfo {author} {\bibfnamefont {Z.}~\bibnamefont {Wang}},
  \bibinfo {author} {\bibfnamefont {K.}~\bibnamefont {Takata}}, \bibinfo
  {author} {\bibfnamefont {R.~L.}\ \bibnamefont {Byer}}, \ and\ \bibinfo
  {author} {\bibfnamefont {Y.}~\bibnamefont {Yamamoto}},\ }\href
  {https://doi.org/10.1038/nphoton.2014.249} {\bibfield  {journal} {\bibinfo
  {journal} {Nature Photonics}\ }\textbf {\bibinfo {volume} {8}},\ \bibinfo
  {pages} {937 EP } (\bibinfo {year} {2014})}\BibitemShut {NoStop}%
\bibitem [{\citenamefont {Pal}\ \emph {et~al.}(2017)\citenamefont {Pal},
  \citenamefont {Tradonsky}, \citenamefont {Chriki}, \citenamefont {Friesem},\
  and\ \citenamefont {Davidson}}]{Pal17}%
  \BibitemOpen
  \bibfield  {author} {\bibinfo {author} {\bibfnamefont {V.}~\bibnamefont
  {Pal}}, \bibinfo {author} {\bibfnamefont {C.}~\bibnamefont {Tradonsky}},
  \bibinfo {author} {\bibfnamefont {R.}~\bibnamefont {Chriki}}, \bibinfo
  {author} {\bibfnamefont {A.~A.}\ \bibnamefont {Friesem}}, \ and\ \bibinfo
  {author} {\bibfnamefont {N.}~\bibnamefont {Davidson}},\ }\href {\doibase
  10.1103/PhysRevLett.119.013902} {\bibfield  {journal} {\bibinfo  {journal}
  {Phys. Rev. Lett.}\ }\textbf {\bibinfo {volume} {119}},\ \bibinfo {pages}
  {013902} (\bibinfo {year} {2017})}\BibitemShut {NoStop}%
\bibitem [{\citenamefont {Mahler}\ \emph
  {et~al.}(2019{\natexlab{b}})\citenamefont {Mahler}, \citenamefont {Pal},
  \citenamefont {Tradonsky}, \citenamefont {Chriki}, \citenamefont {Friesem},\
  and\ \citenamefont {Davidson}}]{Mahler19}%
  \BibitemOpen
  \bibfield  {author} {\bibinfo {author} {\bibfnamefont {S.}~\bibnamefont
  {Mahler}}, \bibinfo {author} {\bibfnamefont {V.}~\bibnamefont {Pal}},
  \bibinfo {author} {\bibfnamefont {C.}~\bibnamefont {Tradonsky}}, \bibinfo
  {author} {\bibfnamefont {R.}~\bibnamefont {Chriki}}, \bibinfo {author}
  {\bibfnamefont {A.~A.}\ \bibnamefont {Friesem}}, \ and\ \bibinfo {author}
  {\bibfnamefont {N.}~\bibnamefont {Davidson}},\ }\href {\doibase
  10.1088/1361-6455/ab3d00} {\bibfield  {journal} {\bibinfo  {journal} {Journal
  of Physics B: Atomic, Molecular and Optical Physics}\ }\textbf {\bibinfo
  {volume} {52}},\ \bibinfo {pages} {205401} (\bibinfo {year}
  {2019}{\natexlab{b}})}\BibitemShut {NoStop}%
\bibitem [{\citenamefont {Tradonsky}\ \emph {et~al.}(2019)\citenamefont
  {Tradonsky}, \citenamefont {Gershenzon}, \citenamefont {Pal}, \citenamefont
  {Chriki}, \citenamefont {Friesem}, \citenamefont {Raz},\ and\ \citenamefont
  {Davidson}}]{Tradonsky18}%
  \BibitemOpen
  \bibfield  {author} {\bibinfo {author} {\bibfnamefont {C.}~\bibnamefont
  {Tradonsky}}, \bibinfo {author} {\bibfnamefont {I.}~\bibnamefont
  {Gershenzon}}, \bibinfo {author} {\bibfnamefont {V.}~\bibnamefont {Pal}},
  \bibinfo {author} {\bibfnamefont {R.}~\bibnamefont {Chriki}}, \bibinfo
  {author} {\bibfnamefont {A.~A.}\ \bibnamefont {Friesem}}, \bibinfo {author}
  {\bibfnamefont {O.}~\bibnamefont {Raz}}, \ and\ \bibinfo {author}
  {\bibfnamefont {N.}~\bibnamefont {Davidson}},\ }\href {\doibase
  10.1126/sciadv.aax4530} {\bibfield  {journal} {\bibinfo  {journal} {Science
  Advances}\ }\textbf {\bibinfo {volume} {5}} (\bibinfo {year} {2019}),\
  10.1126/sciadv.aax4530}\BibitemShut {NoStop}%
\bibitem [{\citenamefont {Eckhouse}\ \emph {et~al.}(2008)\citenamefont
  {Eckhouse}, \citenamefont {Fridman}, \citenamefont {Davidson},\ and\
  \citenamefont {Friesem}}]{Eckhouse08}%
  \BibitemOpen
  \bibfield  {author} {\bibinfo {author} {\bibfnamefont {V.}~\bibnamefont
  {Eckhouse}}, \bibinfo {author} {\bibfnamefont {M.}~\bibnamefont {Fridman}},
  \bibinfo {author} {\bibfnamefont {N.}~\bibnamefont {Davidson}}, \ and\
  \bibinfo {author} {\bibfnamefont {A.~A.}\ \bibnamefont {Friesem}},\ }\href
  {\doibase 10.1103/PhysRevLett.100.024102} {\bibfield  {journal} {\bibinfo
  {journal} {Phys. Rev. Lett.}\ }\textbf {\bibinfo {volume} {100}},\ \bibinfo
  {pages} {024102} (\bibinfo {year} {2008})}\BibitemShut {NoStop}%
\bibitem [{\citenamefont {Zhou}\ \emph {et~al.}(2004)\citenamefont {Zhou},
  \citenamefont {Liu}, \citenamefont {Etson}, \citenamefont {Abranyos},
  \citenamefont {Padilla},\ and\ \citenamefont {Chen}}]{Zhou04}%
  \BibitemOpen
  \bibfield  {author} {\bibinfo {author} {\bibfnamefont {Y.}~\bibnamefont
  {Zhou}}, \bibinfo {author} {\bibfnamefont {L.}~\bibnamefont {Liu}}, \bibinfo
  {author} {\bibfnamefont {C.}~\bibnamefont {Etson}}, \bibinfo {author}
  {\bibfnamefont {Y.}~\bibnamefont {Abranyos}}, \bibinfo {author}
  {\bibfnamefont {A.}~\bibnamefont {Padilla}}, \ and\ \bibinfo {author}
  {\bibfnamefont {Y.~C.}\ \bibnamefont {Chen}},\ }\href {\doibase
  10.1063/1.1699448} {\bibfield  {journal} {\bibinfo  {journal} {Applied
  Physics Letters}\ }\textbf {\bibinfo {volume} {84}},\ \bibinfo {pages} {3025}
  (\bibinfo {year} {2004})}\BibitemShut {NoStop}%
\bibitem [{\citenamefont {Guillot}\ \emph {et~al.}(2011)\citenamefont
  {Guillot}, \citenamefont {Desfarges-Berthelemot}, \citenamefont
  {Kerm\`{e}ne},\ and\ \citenamefont {Barth\'{e}l\'{e}my}}]{Guillot11}%
  \BibitemOpen
  \bibfield  {author} {\bibinfo {author} {\bibfnamefont {J.}~\bibnamefont
  {Guillot}}, \bibinfo {author} {\bibfnamefont {A.}~\bibnamefont
  {Desfarges-Berthelemot}}, \bibinfo {author} {\bibfnamefont {V.}~\bibnamefont
  {Kerm\`{e}ne}}, \ and\ \bibinfo {author} {\bibfnamefont {A.}~\bibnamefont
  {Barth\'{e}l\'{e}my}},\ }\href {\doibase 10.1364/OL.36.002907} {\bibfield
  {journal} {\bibinfo  {journal} {Opt. Lett.}\ }\textbf {\bibinfo {volume}
  {36}},\ \bibinfo {pages} {2907} (\bibinfo {year} {2011})}\BibitemShut
  {NoStop}%
\bibitem [{\citenamefont {Sivaramakrishnan}\ \emph {et~al.}(2015)\citenamefont
  {Sivaramakrishnan}, \citenamefont {Chang}, \citenamefont {Galvanauskas},\
  and\ \citenamefont {Winful}}]{Sivaramakrishnan15}%
  \BibitemOpen
  \bibfield  {author} {\bibinfo {author} {\bibfnamefont {S.}~\bibnamefont
  {Sivaramakrishnan}}, \bibinfo {author} {\bibfnamefont {W.~Z.}\ \bibnamefont
  {Chang}}, \bibinfo {author} {\bibfnamefont {A.}~\bibnamefont {Galvanauskas}},
  \ and\ \bibinfo {author} {\bibfnamefont {H.~G.}\ \bibnamefont {Winful}},\
  }\href {\doibase 10.1109/JQE.2015.2449310} {\bibfield  {journal} {\bibinfo
  {journal} {IEEE Journal of Quantum Electronics}\ }\textbf {\bibinfo {volume}
  {51}},\ \bibinfo {pages} {1} (\bibinfo {year} {2015})}\BibitemShut {NoStop}%
\bibitem [{\citenamefont {Nixon}\ \emph
  {et~al.}(2013{\natexlab{b}})\citenamefont {Nixon}, \citenamefont {Katz},
  \citenamefont {Small}, \citenamefont {Bromberg}, \citenamefont {Friesem},
  \citenamefont {Silberberg},\ and\ \citenamefont {Davidson}}]{Nixon13_2}%
  \BibitemOpen
  \bibfield  {author} {\bibinfo {author} {\bibfnamefont {M.}~\bibnamefont
  {Nixon}}, \bibinfo {author} {\bibfnamefont {O.}~\bibnamefont {Katz}},
  \bibinfo {author} {\bibfnamefont {E.}~\bibnamefont {Small}}, \bibinfo
  {author} {\bibfnamefont {Y.}~\bibnamefont {Bromberg}}, \bibinfo {author}
  {\bibfnamefont {A.~A.}\ \bibnamefont {Friesem}}, \bibinfo {author}
  {\bibfnamefont {Y.}~\bibnamefont {Silberberg}}, \ and\ \bibinfo {author}
  {\bibfnamefont {N.}~\bibnamefont {Davidson}},\ }\href
  {https://doi.org/10.1038/nphoton.2013.248} {\bibfield  {journal} {\bibinfo
  {journal} {Nature Photonics}\ }\textbf {\bibinfo {volume} {7}},\ \bibinfo
  {pages} {919 EP } (\bibinfo {year} {2013}{\natexlab{b}})}\BibitemShut
  {NoStop}%
\bibitem [{\citenamefont {Hercher}(1967)}]{Hercher67}%
  \BibitemOpen
  \bibfield  {author} {\bibinfo {author} {\bibfnamefont {M.}~\bibnamefont
  {Hercher}},\ }\href {\doibase 10.1364/AO.6.000947} {\bibfield  {journal}
  {\bibinfo  {journal} {Appl. Opt.}\ }\textbf {\bibinfo {volume} {6}},\
  \bibinfo {pages} {947} (\bibinfo {year} {1967})}\BibitemShut {NoStop}%
\bibitem [{\citenamefont {Wojcik}\ \emph {et~al.}(2010)\citenamefont {Wojcik},
  \citenamefont {Yu}, \citenamefont {Diehl}, \citenamefont {Capasso},\ and\
  \citenamefont {Belyanin}}]{Wojcik10}%
  \BibitemOpen
  \bibfield  {author} {\bibinfo {author} {\bibfnamefont {A.~K.}\ \bibnamefont
  {Wojcik}}, \bibinfo {author} {\bibfnamefont {N.}~\bibnamefont {Yu}}, \bibinfo
  {author} {\bibfnamefont {L.}~\bibnamefont {Diehl}}, \bibinfo {author}
  {\bibfnamefont {F.}~\bibnamefont {Capasso}}, \ and\ \bibinfo {author}
  {\bibfnamefont {A.~A.}\ \bibnamefont {Belyanin}},\ }\href {\doibase
  10.1117/1.3498773} {\bibfield  {journal} {\bibinfo  {journal} {Optical
  Engineering}\ }\textbf {\bibinfo {volume} {49}},\ \bibinfo {pages} {1 }
  (\bibinfo {year} {2010})}\BibitemShut {NoStop}%
\bibitem [{\citenamefont {Winful}\ and\ \citenamefont
  {Walton}(1992)}]{Winful92}%
  \BibitemOpen
  \bibfield  {author} {\bibinfo {author} {\bibfnamefont {H.~G.}\ \bibnamefont
  {Winful}}\ and\ \bibinfo {author} {\bibfnamefont {D.~T.}\ \bibnamefont
  {Walton}},\ }\href {\doibase 10.1364/OL.17.001688} {\bibfield  {journal}
  {\bibinfo  {journal} {Opt. Lett.}\ }\textbf {\bibinfo {volume} {17}},\
  \bibinfo {pages} {1688} (\bibinfo {year} {1992})}\BibitemShut {NoStop}%
\bibitem [{\citenamefont {Lacot}\ and\ \citenamefont
  {Stoeckel}(1996)}]{Lacot96}%
  \BibitemOpen
  \bibfield  {author} {\bibinfo {author} {\bibfnamefont {E.}~\bibnamefont
  {Lacot}}\ and\ \bibinfo {author} {\bibfnamefont {F.}~\bibnamefont
  {Stoeckel}},\ }\href {\doibase 10.1364/JOSAB.13.002034} {\bibfield  {journal}
  {\bibinfo  {journal} {J. Opt. Soc. Am. B}\ }\textbf {\bibinfo {volume}
  {13}},\ \bibinfo {pages} {2034} (\bibinfo {year} {1996})}\BibitemShut
  {NoStop}%
\bibitem [{\citenamefont {Arnaud}(1969)}]{Arnaud69}%
  \BibitemOpen
  \bibfield  {author} {\bibinfo {author} {\bibfnamefont {J.~A.}\ \bibnamefont
  {Arnaud}},\ }\href {\doibase 10.1364/AO.8.000189} {\bibfield  {journal}
  {\bibinfo  {journal} {Appl. Opt.}\ }\textbf {\bibinfo {volume} {8}},\
  \bibinfo {pages} {189} (\bibinfo {year} {1969})}\BibitemShut {NoStop}%
\bibitem [{\citenamefont {Fox}\ and\ \citenamefont {Li}(1961)}]{FoxLi61}%
  \BibitemOpen
  \bibfield  {author} {\bibinfo {author} {\bibfnamefont {A.~G.}\ \bibnamefont
  {Fox}}\ and\ \bibinfo {author} {\bibfnamefont {T.}~\bibnamefont {Li}},\
  }\href {\doibase 10.1002/j.1538-7305.1961.tb01625.x} {\bibfield  {journal}
  {\bibinfo  {journal} {Bell System Technical Journal}\ }\textbf {\bibinfo
  {volume} {40}},\ \bibinfo {pages} {453} (\bibinfo {year} {1961})}\BibitemShut
  {NoStop}%
\bibitem [{\citenamefont {Gerchberg}\ and\ \citenamefont
  {Saxton}(1972)}]{Gerchberg72}%
  \BibitemOpen
  \bibfield  {author} {\bibinfo {author} {\bibfnamefont {R.~W.}\ \bibnamefont
  {Gerchberg}}\ and\ \bibinfo {author} {\bibfnamefont {W.~O.}\ \bibnamefont
  {Saxton}},\ }\href@noop {} {\bibfield  {journal} {\bibinfo  {journal}
  {Optik}\ }\textbf {\bibinfo {volume} {35}},\ \bibinfo {pages} {237} (\bibinfo
  {year} {1972})}\BibitemShut {NoStop}%
\bibitem [{Not({\natexlab{a}})}]{Note0}%
  \BibitemOpen
  \bibinfo {note} {{T}he in-phase state is the more dominant because it loss is
  slightly smaller than the out-of-phase state (as in
  {F}ig.~\ref{fig:4_better_solver_graphs} (a)).}\BibitemShut {Stop}%
\bibitem [{\citenamefont {Mehuys}\ \emph {et~al.}(1991)\citenamefont {Mehuys},
  \citenamefont {Streifer}, \citenamefont {Waarts},\ and\ \citenamefont
  {Welch}}]{Mehuys91}%
  \BibitemOpen
  \bibfield  {author} {\bibinfo {author} {\bibfnamefont {D.}~\bibnamefont
  {Mehuys}}, \bibinfo {author} {\bibfnamefont {W.}~\bibnamefont {Streifer}},
  \bibinfo {author} {\bibfnamefont {R.~G.}\ \bibnamefont {Waarts}}, \ and\
  \bibinfo {author} {\bibfnamefont {D.~F.}\ \bibnamefont {Welch}},\ }\href
  {\doibase 10.1364/OL.16.000823} {\bibfield  {journal} {\bibinfo  {journal}
  {Opt. Lett.}\ }\textbf {\bibinfo {volume} {16}},\ \bibinfo {pages} {823}
  (\bibinfo {year} {1991})}\BibitemShut {NoStop}%
\bibitem [{Not({\natexlab{b}})}]{Note1}%
  \BibitemOpen
  \bibinfo {note} {{T}o keep our system simple with only two low loss phase
  locked states, we suppressed two additional low loss phase locked
  states.}\BibitemShut {Stop}%
\bibitem [{Not({\natexlab{c}})}]{Note2}%
  \BibitemOpen
  \bibinfo {note} {{T}he eigenvalues calculation uses saturation function for
  the {SA} that contains four parameters ({SA}'s intensities and transmissions
  thresholds) that are adjusted to set the degeneracy distance to agree with
  the {T}albot distance where the out-of-phase state intersects the in-phase
  state in {F}igs.~\ref{fig:4_better_solver_graphs} (b) and (c) that is about
  $1.04Z_{T}$.}\BibitemShut {Stop}%
\bibitem [{Not({\natexlab{d}})}]{Note3}%
  \BibitemOpen
  \bibinfo {note} {{W}e analyzed the far-field intensity distribution of each
  realization and set an intensity threshold value ($\approx15\%$ of the
  maximal intensity) for determining whether the in-phase or out-of-phase or
  coexistence state is present.}\BibitemShut {Stop}%
\bibitem [{\citenamefont {Stanley}(1971)}]{Stanley71}%
  \BibitemOpen
  \bibfield  {author} {\bibinfo {author} {\bibfnamefont {H.~E.}\ \bibnamefont
  {Stanley}},\ }\href@noop {} {\emph {\bibinfo {title} {Introduction to phase
  transitions and critical phenomena}}}\ (\bibinfo  {publisher} {Oxford
  University Press},\ \bibinfo {address} {New York},\ \bibinfo {year}
  {1971})\BibitemShut {NoStop}%
\bibitem [{\citenamefont {Strecka}\ and\ \citenamefont
  {Jascur}(2015)}]{Strecka05}%
  \BibitemOpen
  \bibfield  {author} {\bibinfo {author} {\bibfnamefont {J.}~\bibnamefont
  {Strecka}}\ and\ \bibinfo {author} {\bibfnamefont {M.}~\bibnamefont
  {Jascur}},\ }\href@noop {} {\bibfield  {journal} {\bibinfo  {journal} {Acta
  Physica Slovac}\ }\textbf {\bibinfo {volume} {65}},\ \bibinfo {pages} {235}
  (\bibinfo {year} {2015})}\BibitemShut {NoStop}%
\end{thebibliography}%
\end{document}